\documentclass[a4paper,11pt]{article}

\usepackage{amssymb}
\usepackage{amsmath}
\usepackage{graphicx}
\usepackage{comment}

\begin{document}

\newcommand\niv{\ensuremath{\gamma J}}
\newcommand{\unit}[2]{\ensuremath{#1\;\mathrm{#2}}}
\newcommand{\proba}{\ensuremath{\mathcal{P}}}
\newcommand{\partit}{\ensuremath{\mathcal{U}}}

\huge

\begin{center}
A consistent approach for mixed detailed and statistical calculation of opacities in hot plasmas
\end{center}

\vspace{1cm}

\large

\begin{center}
Quentin Porcherot$^1$, Jean-Christophe Pain$^1$, Franck Gilleron$^1$ and Thomas Blenski$^2$
\end{center}

\vspace{1cm}

\normalsize

\begin{center}
$^1$CEA, DAM, DIF, F-91297 Arpajon, France\\
$^2$CEA, DSM, IRAMIS, F-91191 Gif-sur-Yvette, France
\end{center}
\begin{abstract}
Absorption and emission spectra of plasmas with multicharged-ions contain transition arrays with a huge number of coalescent electric-dipole (E1) lines, which are well suited for treatment by the unresolved transition array and derivative methods. But, some transition arrays show detailed features whose description requires diagonalization of the Hamiltonian matrix. We developed a hybrid opacity code, called SCORCG, which combines statistical approaches with fine-structure calculations consistently. Data required for the computation of detailed transition arrays (atomic configurations and atomic radial integrals) are calculated by the super-configuration code SCO (\emph{Super-Configuration Opacity}), which provides an accurate description of the plasma screening effects on the wave-functions. Level energies as well as position and strength of spectral lines are computed by an adapted RCG routine of R. D. Cowan. The resulting code provides opacities for hot plasmas and can handle mid-$Z$ elements. The code is also a powerful tool for the interpretation of recent laser and \emph{Z}-pinch experimental spectra, as well as for validation of statistical methods.
\end{abstract}

\section{Introduction}
\label{intro}

Computation of plasma opacities is necessary in many fields of plasma physics: experiments are diagnosed with emission or transmission spectra and mean opacities are required to evaluate radiative transfer.
Both applications require calculation of spectra with a {\em high} resolving power, i.e., $\frac{h\nu} {\Delta h\nu}$ of over a few hundreds, which involves computation of atomic structure and transition arrays.
However, plasma spectroscopy rarely requires one to get over a resolving power higher than a few thousands, thus fine structure modeling provides sufficiently accurate spectra.

Nowadays, plasma opacity is calculated by codes which use DLA (Detailed Line Accounting) like HULLAC \cite{HULLAC2001} and FAC \cite{Gu2008} or, for faster computation, statistical methods based on UTA (Unresolved Transition Array) or STA (Super Transition Array) approaches.

DLA approach gives precise enough transition rates and energies for spectroscopy and mean opacities calculation, but requires substantial resources as soon as configurations get more complex (degeneracy $g_C \gtrsim 1000$).
However, the DLA approach alone enables one to account for the effects of the Boltzmann factor $e^{-\beta E_{\niv}}$ on probability of levels \niv{} within a configuration, even if statistical approaches can still be improved [Gilleron et al., this issue].
The DLA approach is also useful when transition arrays have small total physical broadening, due to various mechanisms, e.g., Doppler, Stark, collisions.

On the other hand, statistical approaches enable much faster opacity calculations, because order 0 to 2 moments of transition arrays can be obtained without diagonalization of the Hamiltonian matrix.
That is the basis of UTA (non-relativistic) \cite{BauchesI} and SOSA (relativistic Spin-Orbit Split Array) \cite{BauchesIII} approaches.
UTA and SOSA still require detailed configuration accounting (DCA).
In case of a large number of configurations, the STA approach \cite{STA} enables an even faster calculation of plasma opacities, replacing configurations by superconfigurations.
SCO \cite{Blenski2k} is able to handle both STA and DCA for opacity calculation.
However, statistical approaches are limited by the spectral accuracy they can reach, and, thus, are not able to resolve all structures experimental spectra can show. Further, statistical approaches neglect the effects of temperature on probability distribution over levels of a single configuration. This probability distribution can have a strong impact on transition array shape if the plasma temperature is of the same order of magnitude as the configuration energy span, or {\em receptive zone}.

However, it is worth investigating the relevance of a DLA computation of very complex transition arrays.
As shown in the examples in table \ref{tab:complexity}, complex transition arrays with millions of lines can occur in mid-$Z$ plasmas at moderate temperatures, e.g., copper at 20 eV, for doubly-excited configurations, when the $3d$ subshell is almost half-filled.
In cases when a $f$ subshell is partially opened, even singly-excited configurations show a very complex structure.
Since statistical and detailed approaches are both useful for an accurate description of spectra over a broad range of temperature, density and atomic number, a model mixing both approaches could substantially improve the quality of the opacity calculation, without requiring too many computational resources.

\begin{table}
  \centering
  \begin{tabular}[c]{|l|c|c|}\hline
    Ion & Transition array & Number of lines \\ \hline
    Fe XIV at 150 eV & \normalsize{$2p^5 3s^2 3p 3d - 2p^4 3s^2 3p 3d^2$} & 177,684 \\
    Cu IX at 40 eV & $2p^6 3p^4 3d^5-2p^5 3p^4 3d^6$ & 564,293  \\ 
    Gd VI at 40 eV & $4f^4 5d - 4f^3 5d 6d$ & 1,139,911 \\ \hline
  \end{tabular}
  \caption{Number of spectral lines of transition arrays starting from singly- or doubly-excited configurations.}
  \label{tab:complexity}
\end{table} 

\section{The hybrid approach}
The hybrid approach we use in our model should be able to compute accurate spectra within a reasonable time, mixing configurations and superconfigurations at local thermodynamic equilibrium (LTE).
Bound-bound opacity may include detailed transition arrays, as well as UTA, SOSA and STA.
The calculation should be as consistent as possible from a physical point of view.
Effects of the plasma environment on atomic structure should be considered even in DLA computations.
We also intend to provide an automatic tool, able to select the (super)configurations to be studied as well as to decide how to handle each transition array.

\subsection{Implementation}
We choose to use two existing codes, SCO and R. D. Cowan's RCG routine \cite{TASS}.
SCO is a superconfiguration opacity code based on a physical model that gives a reliable description of plasma physics over a wide range of temperature, density and atomic number.
A special feature of SCO is the automatic selection of configurations and superconfigurations, starting from an average atom model and using fluctuation theory \cite{perrot88,wilson93}.
Atomic structure is computed using a self-consistent approach within central-field approximation, and density effects like free-electron screening are taken into account by the code.
Orbital relaxation can be included for selected transitions.
There are relativistic corrections based on Pauli approximation, which enables correct handling of medium to high $Z$ elements.

On the other hand, Cowan's code can compute detailed transition arrays, given an ion and its electronic configuration.
Atomic structure and atomic spectra are calculated by two different programs, respectively RCN/RCN2 and RCG.
Output of the former is automatically formatted as input (atomic structure integrals) for the latter.
The separation between the calculations of atomic structure and of transition arrays enables the input of atomic structure integrals computed by other processes, e.g., SCO.
We choose to adapt RCG as a subroutine in SCO only used for DLA computation.

\subsection{Description of SCORCG}
The resulting code is able to handle consistently both detailed and statistical transition arrays.
Input is a Fortran 77 list of variables in which thermodynamic quantities, a list of subshells needed, the number of superconfigurations and the ionization range, among other options, can be set.
Most parameters have a default setting suitable for standard calculations, thus input can be rather small.
An average atom plasma is computed at first.
It enables one to have an overlook at the atomic structure and other global properties of the plasma to be studied.
For the computation itself, this first run selects the configurations that are to be included in the opacity calculation.
The code can be stopped after this first run and the list of selected configurations can be kept as is or modified.

In SCORCG, a superconfiguration $\Xi$ can match a single configuration.
Such single superconfigurations are treated with the DCA methods.
Other superconfigurations are specifically called non-DCA superconfigurations.
For each superconfiguration $\Xi$ in the list generated as input, SCO processes self-consistent calculations of atomic structure, with relativistic corrections, free-electron screening inside ions.
Effects of neighbouring ions are limited by the Wigner-Seitz sphere.
Orbital relaxation can be included if wished \cite{perrot97relax}.
Next, SCO computes all transition arrays starting from $\Xi$.

Only transition arrays starting from single configurations can be detailed.
SCO handles all cases of UTA, SOSA and STA, and sends to the modified RCG routine arrays that are to be treated in detail, according to criteria driven by the physical broadening as well as computational issues.
The spectral opacity of the elemental plasma can be written as the sum of the opacities of the species in it, weighted by their respective probabilities:
\begin{equation}
  \label{eq:opac}
  \kappa (h\nu) = \sum_{\Xi} \proba(\Xi) \kappa_{\Xi} (h\nu),
\end{equation}
where $\Xi$ is any superconfiguration in the plasma, $\proba(\Xi)$ its probability, and $\kappa_{\Xi} (h\nu)$ its spectral opacity was $\Xi$ the only species in the plasma. % références ?

The modified RCG routine does not only compute line strengths and energies with radial integrals given by SCO for the detailed configuration $C$, it also calculates other features of the detailed transition arrays, such as strength-weighted moments of the distribution of lines.
Moments of orders up to 100 can be calculated in statistical weight approximation (SWAP) like in Eqs. (\ref{eq:momc}) and (\ref{eq:momr}) and by using the Boltzmann factor. The detailed partition function $\partit_C$ of the starting configuration is also calculated according to Eq. (\ref{Eq:UC}) by our modified RCG routine. Each transition array starting from $C$ is then multiplied by $\partit_C$.

\subsection{Management of different types of species in plasmas}
% We propose the splitting of the subsection, as it was much lengthened

The general expression for the partition function of the configuration $C$ is
\begin{equation}
  \label{Eq:UC}
  \partit_C = \sum_{\niv \in C} (2J+1) e^{-\beta E_{\niv}},
\end{equation}%
where $\niv$ is a fine-structure level of $C$, characterized by its total angular momentum $J$ and its energy $E_{\niv}$; $\beta$ is the inverse of temperature $kT$.
This partition function is used in SCORCG when at least one transition array starting from the configuration is calculated in detail.

If all transition arrays starting from $C$ are statistically treated, then $C$ will not be detailed.
Statistical treatment of configurations and transition arrays assumes the probability of each level $\niv$ does not depend on temperature.
Therefore, the partition function of $C$ only depends on the population $w_s$ of the subshell $s$ characterized by its angular momentum $\ell_s$: 
\begin{equation}
  \partit_C = \sum_{\niv \in C} (2J+1) e^{-\beta E_C} = g_C e^{-\beta E_C},
\end{equation}
with 
\begin{equation}
  g_C = \sum_{\niv \in C} (2J+1) = \prod_{s}\left(
  \begin{array}{c}
    4 \ell_s + 2 \\ w_s
  \end{array}\right)
\label{eq:dege}.
\end{equation}

After the computation of the opacities of all superconfigurations, the total bound-free and bound-bound opacities are divided by the total partition function $\partit$ of the plasma:
\begin{equation}
  \partit = \sum_{C\:\mathrm{detailed}} \partit_C + \sum_{C\:\mathrm{statistical}} \partit_C + \sum_{\Xi\:\mathrm{non-DCA}} \partit_{\Xi}.
\end{equation}

Hence, the opacity of each superconfiguration $\Xi$ in Eq. (\ref{eq:opac}) is multiplied by the probability of the latter, was $\Xi$ detailed, DCA or non-DCA:
\begin{equation}
  \proba(\Xi)=\frac{\partit_{\Xi}}{\partit}.
\end{equation}

\subsection{Current limitations of the model}
There are two types of limitations in SCORCG: the first are computational while the second are due to the model.
Little computational resources are required by a STA or UTA code, and one can run such opacity programs on a desktop computer.
DLA greatly increases the complexity and slows down computation if the detailed configurations are complex.

The computational limitations are propagated to the generation of the opacity from our modified RCG routine. Hence the DLA approach is limited to transition arrays with less than 800,000 lines, a limit that can be set lower upon input file.
For one transition array, the size of blocks with the same $J$ in Hamiltonian matrix is limited to 4,000. However, this limitation is much less restrictive than the restriction of lines in a transition array.
Moreover, a maximum of 8 subshells can be opened in both configurations.

The plasma is modeled using average-atom and (super)configurations.
Fine-structure information is only used for calculation of lines or partition functions; atomic structure and modeling of plasma density effects are statistically treated.
Relativistic effects are accounted in the Pauli approximation, which provides accurate results for mid-$Z$ elements.
Full configuration interaction (CI) is not implemented in SCORCG.
Only configuration interactions between relativistic subconfigurations that belong to the same configuration are handled in the calculation (exactly in DLA part and by an approximate method in the statistical part).

DLA calculations are much more time-consuming than UTA or STA.
For most transition arrays, a great deal of computational resource is consumed in the convolution of transitions with physical line profiles.
The first solution we implemented is to regroup all lines of a transition array into energy bins in which spectral line intensities are added.
In SCORCG, we assume that all lines of the same transition array have the same broadening as the statistical transition array for which SCO routines compute Stark and collisional formulas \cite{Dimitrijevic87}.
Even if ``binning'' reduces the amount of resources required for a single transition array with many lines, numerous lines remain to be computed within a whole opacity calculation.
For further improvement of DLA efficiency, one can use one average broadening per bin for all transition arrays.

Abdallah et al. \cite{abdallah2007} showed this approach gives moderately accurate spectra, in that Rosseland mean opacity changed by about 1 or 2 \% from a calculation with a detailed line by line convolution.
However, the method they developed for two broadening bins per energy bin -- one for narrow lines and one for wide lines -- gives much more accurate results, with almost no observable difference for the opacity spectrum and yielding a Rosseland error of about 0.01 \%.
The latter method requires one to sample all spectral lines twice: first for computation of average broadenings for each energy bin, and second for filling bins with spectral lines that fit the criteria of energy and broadening.

\section{Experiment interpretations}
\label{sec:experiment}
We have analyzed experimental spectra produced in laser and $Z$-pinch facilities, focusing on spectra for which there is no satisfactory interpretation done when statistical codes were used.
Three spectra for which our hybrid model strongly improves the agreement with experimental data are presented. The first spectrum was recorded in an experiment at the HELEN laser facility in the United Kingdom \cite{Davidson87}.
The second was obtained in an experiment at Sandia $Z$-pinch facility in the USA\cite{Bailey2k7}.
The third, most recent, is a transmission spectrum recorded at LULI 2000 laser facility of Ecole Polytechnique in France \cite{Loisel2k9}.

\subsection{Aluminum on HELEN laser facility}
\label{sec:Al_HELEN}
The aluminum experiment on HELEN laser facility was thought to be at a temperature of \unit{40}{eV} and a density of \unit{0.01}{g/cc} \cite{Davidson87}.
The experimental transmission shows spectrally-resolved features that were not well reproduced by any statistical approach.
Interpretation of this spectrum using SCORCG gives a much better agreement, with a temperature of approximately 40 eV if density effects are not accounted, and of 36 eV if these effects are added.
Although not understood, the differences we observe around \unit{1530}{eV} cannot be explained by temperature and density gradients in the plasma.
\begin{figure}[htp]
  \centering
  \includegraphics[width=\textwidth]{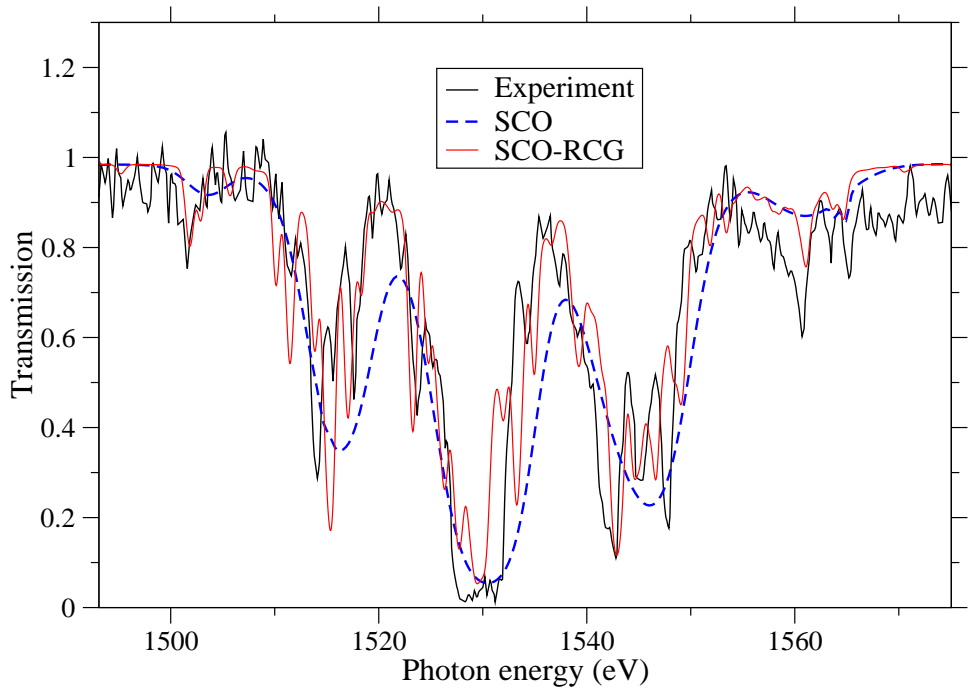}
  \caption{Aluminum spectrum from the experiment of Davidson \cite{Davidson87} compared with SCO and SCORCG calculations.}
  \label{fig:AlDavidson}
\end{figure}

\subsection{Iron on $Z$-pinch}
\label{sec:FeSandia}
The iron spectrum produced at Sandia National Laboratory (SNL) \cite{Bailey2k7} has a high resolution and shows very detailed features.
This transmission spectrum was intended to provide a benchmark spectrum which could test opacity simulations at a temperatures above 100 eV and electron densities higher than \unit{10^{22}}{cm^{-3}}.
\begin{figure}[htp]
  \includegraphics[width=\textwidth]{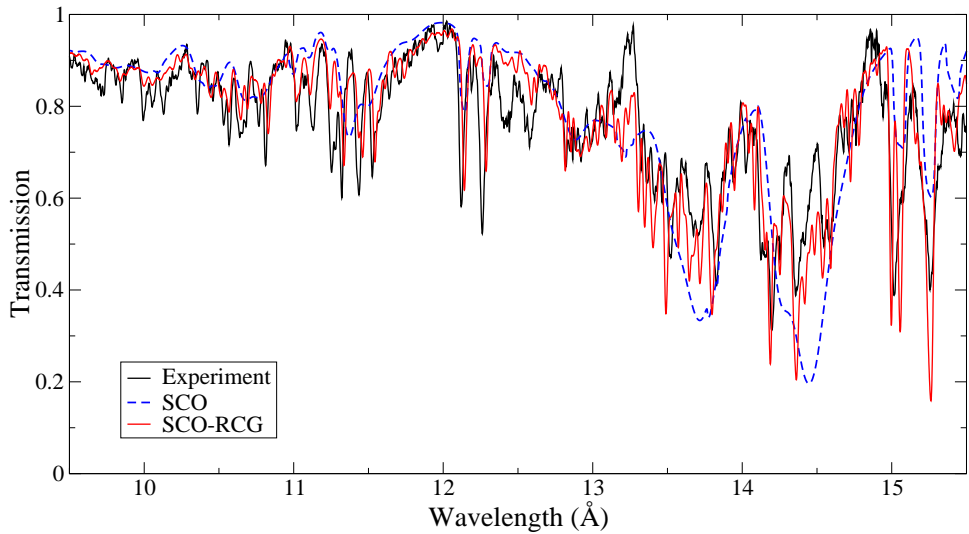}
  \caption{Iron spectrum of Bailey \cite{Bailey2k7}
 and the best agreement found with SCO and SCORCG.}
  \label{fig:FeBailey}
\end{figure}
This spectrum shows features no statistical approach is able to reproduce.
Only codes using a DLA approach can obtain good agreement with the experimental spectrum, and the results already obtained by other codes , e.g., PrismSPECT \cite{PrismSPECT} make interpretation challenging.
The best agreement for the SCORCG calculation with experiment is at $T=\unit{150}{eV}$ and $\rho=\unit{58}{mg/cm^3}$.
Some features do not seem to be reproduced by our mixed detailed and statistical approach, but ion by ion investigations show the discrepancies do not come from ionic distribution.

\subsection{Copper on LULI 2000 laser facility}
\label{sec:CuLULI}
For the LULI experiments we mainly focus on shot 31 of 2008 campaign for opacities of mid-$Z$ elements \cite{Loisel2k9}.
Shot 31 had a copper sample heated to \unit{16}{eV} and with a density of \unit{5}{mg/cc}.
The larger areal mass (\unit{40}{\mu g/cm^2}) makes this transmission spectrum more difficult to model, as bound-bound structures are prevalent.
\begin{figure}[htp]
  \centering
  \includegraphics[width=\textwidth]{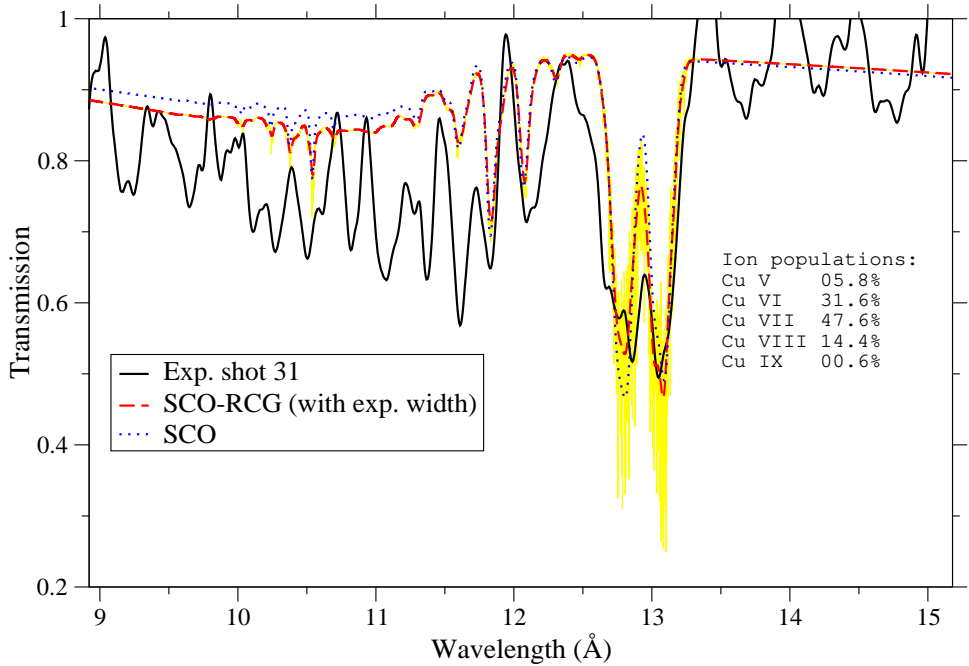}
  \caption{Transmission spectrum of copper obtained in shot 31 \cite{Loisel2k9} and theoretical spectra computed by SCO and SCORCG.}
  \label{fig:CuLULI}
\end{figure}

Even if SCO gives good agreement, the ratio between relativistic substructures $2p_{1/2} - 3d_{3/2}$ and $2p_{3/2} - 3d_{5/2}$ at 12.8 and 13.1 Angstroms is not reproduced by the fully statistical model.
But the SCORCG transmission, like other DLA codes [Blenski et al., this issue], gives good agreement with the experiment.
It is an effect of relativistic configuration interaction between the two major SOSAs of $2p - 3d$ transition.
However, a new relativistic CI model \cite{Gilleron2007} shows a much better agreement of SCO with experiment and DLA codes.

\section{Array shapes and moments}

With its ability to compute transition arrays in both statistical and DLA approaches, SCORCG allows us to perform detailed investigations of the first to fourth order moments of the transition arrays and their consequences for the spectra.

\subsection{Effects of order 3 and 4 moments on modeling of transition arrays}
\label{sec:mom34}
\begin{figure}[htp]
  \centering
  \includegraphics[width=\textwidth]{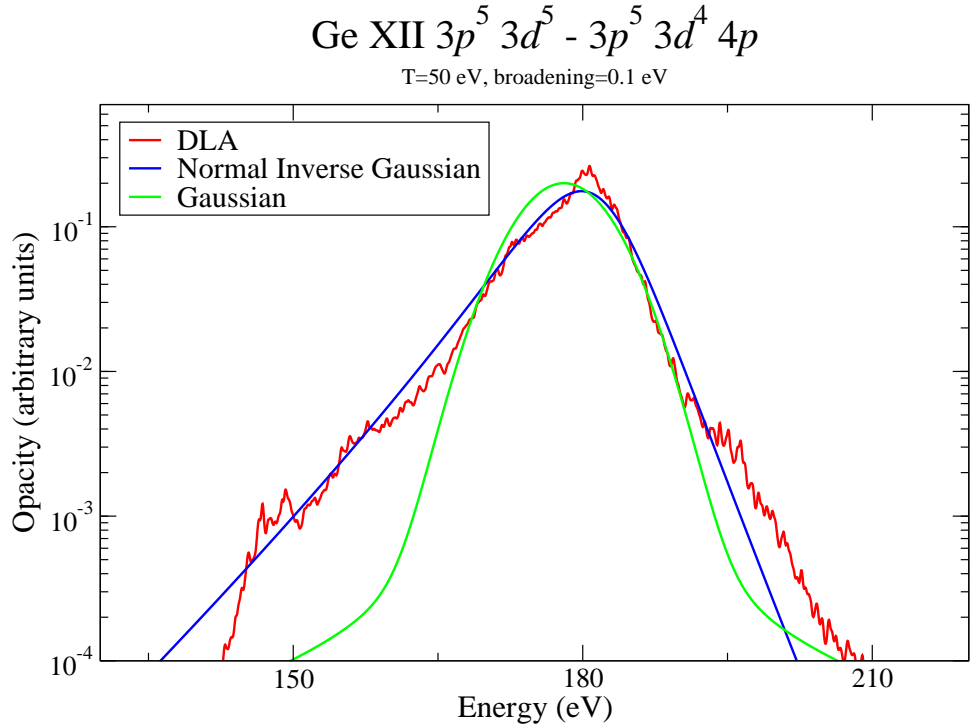}
  \caption{Transition array $3p^5 3d^5 - 3p^4 3d^5 4p$ from Ge XII with broadening of 0.1 eV and temperature of 50 eV. Comparison DLA/UTA-Gaussian/NIG.}
  \label{fig:GeXII}
\end{figure}

Line distribution in transition arrays is modeled by a Gaussian distribution in any UTA approach.
However, if spin-orbit splitting effects are significant, the array is split into several subarrays.
Even when spin-orbit splitting is very low, the distribution of lines in the transition arrays is often not symmetric and neither the central peak nor the wings of the transition array are well-fit by a Gaussian, as shown in Fig. \ref{fig:GeXII}.
We define the order $n$ moment of a transition array with this formula:
\begin{equation}
  \label{eq:mom}
  \mu_{n} = \frac{\displaystyle{\sum_{\gamma J - \gamma ' J'} E_{\gamma J - \gamma ' J'}^n\cdot \mathcal{S}_{\gamma J - \gamma ' J'}}} {\displaystyle\sum_{\gamma J - \gamma ' J'} \mathcal{S}_{\gamma J - \gamma ' J'}}.
\end{equation}
$\displaystyle\sum_{\gamma J - \gamma ' J'} \mathcal{S}_{\gamma J - \gamma ' J'}$ is the area of the transition array and $\mu_1$ is the average energy.
For order 2 and higher, we define centered moments as:
\begin{equation}
  \label{eq:momc}
  \mu_{n}^{(c)} = \frac{\displaystyle{\sum_{\gamma J - \gamma ' J'}(E_{\gamma J - \gamma ' J'}-\mu_1)^n\cdot \mathcal{S}_{\gamma J - \gamma ' J'}}} {\displaystyle{\sum_{\gamma J - \gamma ' J'} \mathcal{S}_{\gamma J - \gamma ' J'}}}.
\end{equation}
$\mu_2^{(c)}$ is the variance of the transition array.
For higher orders, we use the reduced centered moments of the transition array, i.e.,
\begin{equation}
  \label{eq:momr}
  \alpha_n = \frac{\mu_{n}^{(c)}}{\left(\mu_{2}^{(c)}\right)^{\frac{n}{2}}}.
\end{equation}
Order 3 moment is the \emph{skewness} of the array, which represents the asymmetry of the line distribution.
Order 4 moment is the \emph{kurtosis} of the array, which represents the sharpness/flattening of the distribution near its center.
Pain et al. \cite{PG2k9m34} demonstrated how third and fourth orders moments can significantly improve the modeling of transition arrays in a statistical approach.
Figure \ref{fig:GeXII} shows that inclusion of the asymmetry and kurtosis improves the modeling of transition arrays with low spin-orbit splitting.
The normal inverse Gaussian, NIG, agrees much better with the shape of the wings of the transition array, and this improved profile can have an impact on Rosseland mean opacity calculated by statistical approaches.

\begin{figure}[htp]
  \centering
  \includegraphics[width=\textwidth]{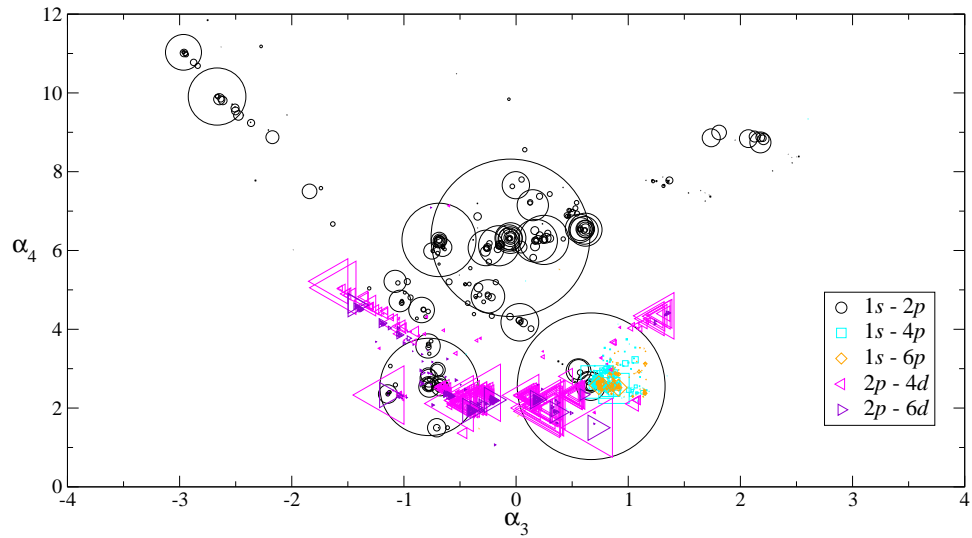}
  \caption{Distribution of order 3 and 4 moments used in the interpretation of Davidson's aluminum experiment. Areas enclosed by circles, squares and triangles are proportional to the product of the total transition array intensity and the abundance of the initial configuration.}
  \label{fig:mom34}
\end{figure}

Figure \ref{fig:mom34} shows that most transition arrays, even those that contribute most to the opacity, are asymmetrical and have a broad range of values for the kurtosis.
Transition arrays seem to converge toward non-Gaussian shapes when $n$ grows.
Those starting from subshell $1s$ tend to concentrate near $(\alpha_3,\alpha_4)=(+0.8, 2.5)$.

Accounting for the order 3 and 4 moments can strongly improve the statistical modeling of some transition arrays with a very peaked shape.
These moments can be obtained by a DLA calculation, knowing that for transition arrays with a large number of lines, a large part of computation time is spent in convolving all lines by Voigt profiles.

\subsection{Effects of temperature on average energy and variance}
\label{sec:mom12}
Temperature, or more exactly the inverse of temperature $\beta = \frac{1}{kT}$ has a significant effect on shape of transition arrays.
Gilleron et al. in Ref. [this issue] show significant impact of temperature on the total absorption of a transition array, even at temperatures higher than the statistical width of the transition array.
Investigations into effects of temperature on the average energy and dispersion of the transition array show that the average energy does not seem to depend much on temperature, because all lines are in a limited energy range.
But the spectral resolution required for an accurate spectral calculation is about 1000, whereas temperature impact on average energy of the transition array is about 10~\% for some transition arrays.
\begin{figure}[htp]
  \centering
  \includegraphics[width=\textwidth]{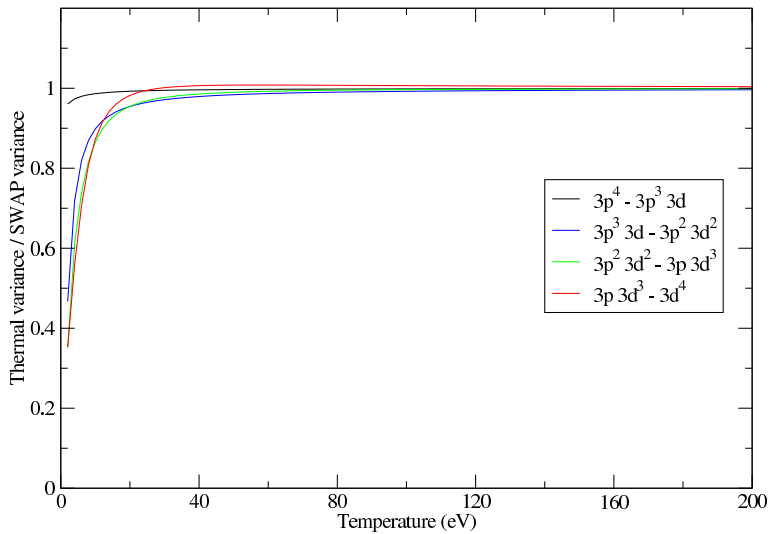}
  \caption{Variances of selected transition arrays in Fe XII.}
 \label{fig:FeMom}
\end{figure}

Figure \ref{fig:FeMom} shows that dependence is quite strong for variance, which can be partly explained by the fact that at low temperature, only transitions starting from ground level of the initial configuration are significant.
This lowers the dispersion of the transition array.
The moments of the transition array $3p^4$ - $3p^3 3d$, of type $\ell^{w+1} - \ell^w \ell'$, are not much affected by temperature.
However, the moments of transition arrays starting from configurations with two open subshells are significantly affected by low temperature.
This can be explained by a wider span in energy for the starting configuration.

\subsection{Effects of the maximum number of lines per transition array on mean opacities}
\label{sec:mean}
\begin{figure}[ht]
  \centering
  \includegraphics[width=\textwidth]{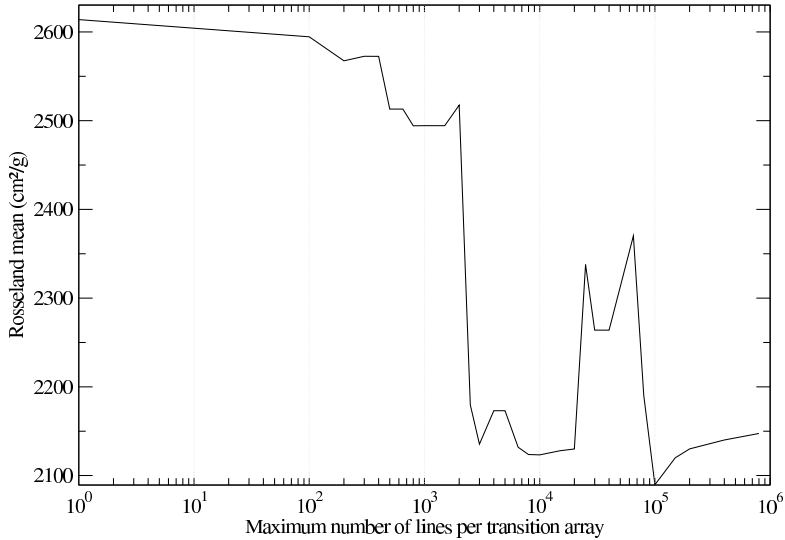}
  \caption{Rosseland mean opacity plotted against the maximum number of lines per transition array}
  \label{fig:CuRosseland}
\end{figure}

As a harmonic mean, the Rosseland mean opacity is sensitive to the presence and the treatment of spectral lines when physical broadening does not merge them into smoother structures.
With our new hybrid code, we started to investigate the effect on the Rosseland mean opacity of changing the DLA treatment.
We focused on one interesting case: copper at \unit{20}{eV} and \unit{10^{-4}}{g/cc}. At this temperature, the Rosseland mean opacity is extremely sensitive to the amount of DLA treatment of the transition arrays.
Between a calculation with as much DLA as possible and a fully statistical calculation, the Rosseland mean differs by 20~\%, as shows Fig. \ref{fig:CuRosseland}.

Figure \ref{fig:CuRosseland} shows that the Rosseland mean opacity only converges to $\kappa_R(\infty)=\unit{2.135}{ cm^2/g}$, the limit of mean opacity with every transition array detailed, if transition arrays with more than 100,000 lines are treated in detail.
Although $\kappa_R(\infty)$ seems to be reached below 10,000 lines per transition array, some structures of the bound-bound spectrum, especially in the $3p-3d$ region at around 60 eV are not converged at this number of lines per array.

\begin{figure}[htp]
  \centering
  \includegraphics[width=\textwidth]{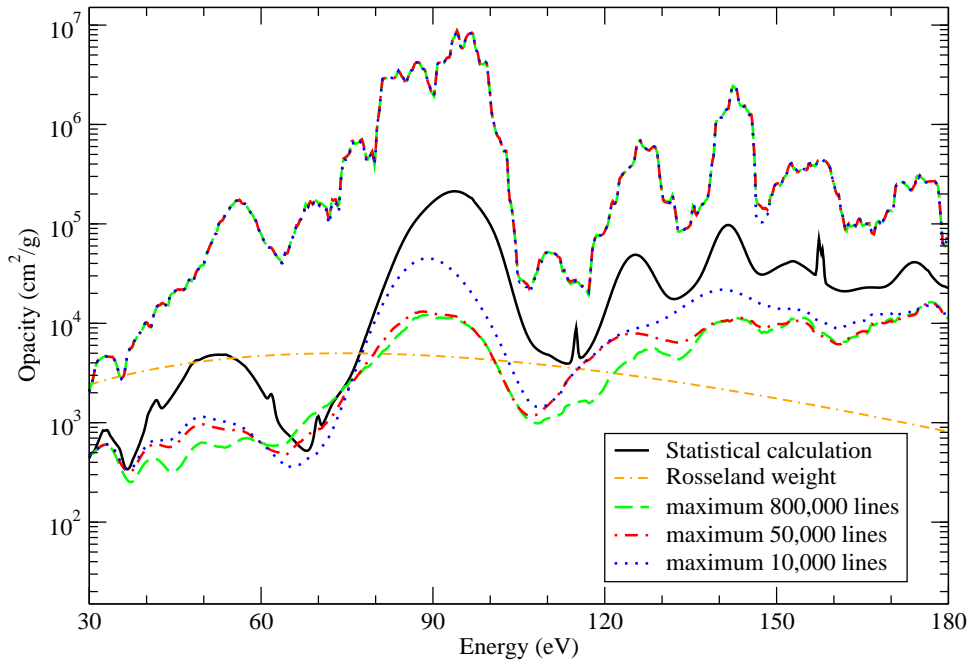}
  \caption{Envelopes of spectra of plasma opacities with different maximum number of lines (upper bounds merge).}
  \label{fig:CuRossSpec}
\end{figure}
Spectral envelopes in Fig. \ref{fig:CuRossSpec} explain the peaks between 25,000 and 65,000 lines per transition array.
In comparison, a similar test on an aluminum plasma shows that convergence of Rosseland mean opacity occurs at a few hundreds of lines per transition array.

\section{Conclusion}
SCORCG, the hybrid code we are developing, is a versatile tool for investigations on plasma opacity computation.
Interpretation of experimental spectra has given results that are comparable to those obtained with other DLA codes.
Beyond this first application, SCORCG enables us to study more deeply statistical properties of bound-bound transitions in hot plasmas by the collection of substantial data, which allows us to automate the execution and produce spectra in a reasonable time.
Therefore, SCORCG will be quite useful in a future investigation into the validity of statistical approximations for opacities in hot plasmas, as it enables one to see their effects on the entire spectrum.
However, this code is still under development, and further improvements involve spectral line handling, criteria for discrimination between transition arrays that require DLA calculation or accounting of order 3 and 4 moments.


\begin{thebibliography}{99}
\bibitem{HULLAC2001} A. Bar-Shalom, M. Klapisch, and J. Oreg, J. Quant. Spectroc. Radiat. Transfer {\bf 71}, 169 (2001). 
\bibitem{Gu2008} M. F. Gu, Can. J. Phys. {\bf 86}, 675 (2008).
\bibitem{BauchesI} C. Bauche-Arnoult, J. Bauche, and M. Klapisch, Phys. Rev. A {\bf 20}, 2424 (1979). 
\bibitem{BauchesIII} C. Bauche-Arnoult, J. Bauche, and M. Klapisch, Phys. Rev. A {\bf 31}, 2248 (1985). 
\bibitem{STA} A. Bar-Shalom, J. Oreg, W. H. Goldstein, D. Shvarts, and A. Zigler, Phys. Rev. A {\bf 40}, 3183 (1989). 
\bibitem{Blenski2k} T. Blenski, A. Grimaldi, and F. Perrot, J. Quant. Spectrosc. Radiat. Transfer {\bf 65}, 91 (2000). 
\bibitem{TASS} R. D. Cowan, The Theory of Atomic Structure and Spectra, The Regents of the University of California, 1981.
\bibitem{perrot88} F. Perrot, Physica A {\bf 150}, 357 (1988).
\bibitem{wilson93} B. G. Wilson, J. Quant. Spectrosc. Radiat. Transfer {\bf 49}, 241 (1993).
\bibitem{perrot97relax} F. Perrot, T. Blenski, and A. Grimaldi, J. Quant. Spectroc. Radiat. Transfer {\bf 58}, 845 (1997). 
\bibitem{Dimitrijevic87} M. S. Dimitrijevic and N. Konjevic, Astron. Astrophys. {\bf 172}, 345 (1987). 
\bibitem{abdallah2007} J. Abdallah. Jr., D. Kilcrease, N. Magee, S. Mazevet, P. Hakel, and M. Sherrill, High Energy Density Phys. {\bf 3}, 309 (2007). 
\bibitem{Davidson87} S. J. Davidson, J. M. Foster, C. C. Smith, K. A. Warburton, and S. J. Rose, App. Phys. Lett. {\bf 52}, 847 (1987). 
\bibitem{Bailey2k7} J. E. Bailey, G. A. Rochau, C. A. Iglesias, J. Abdallah, J. J. MacFarlane, I. Golovkin, P. Wang, R. C. Mancini, P. W. Lake, T. C. Moore, M. Bump, O. Garcia, and S. Mazevet, Phys. Rev. Lett. {\bf 99}, 265002 (2007). 
\bibitem{Loisel2k9} G. Loisel, P. Arnault, S. Bastiani-Ceccotti, T. Blenski, T. Caillaud, J. Fariaut, W. Flsner, F. Gilleron, J.-C. Pain, M. Poirier, C. Reverdin, V. Silvert, F. Thais, S. Turck- Chi\`eze, and B. Villette, High Energy Density Phys. {\bf 5}, 173 (2009). 
\bibitem{PrismSPECT} J. J. Mcfarlane, I. Golovkin, P. Woodruff, and P. Wang, APS Meeting Abstracts {\bf J}, 1181 (2003). 
\bibitem{Gilleron2007} F. Gilleron, J. Bauche, and C. Bauche-Arnoult, J. Phys. B: At. Mol. Opt. Phys. {\bf 40} (2007). 
\bibitem{PG2k9m34} J.-Ch. Pain, F. Gilleron, J. Bauche, and C. Bauche-Arnoult, High Energy Density Phys. {\bf 5}, 294 (2009).
\end{thebibliography}
\end{document}